\documentclass[11pt,notoc]{JHEP3}
\usepackage{epsfig}

\usepackage{mathrsfs}

 \usepackage{graphicx}
 \usepackage{amssymb}
 \usepackage{amsmath}
 \usepackage{amsfonts}
 \usepackage{revsymb}

 \newcommand{\be}{\begin{equation}}
\newcommand{\ee}{\end{equation}}
\newcommand{\bea}{\begin{eqnarray}}
\newcommand{\eea}{\end{eqnarray}}
\newcommand{\bean}{\begin{eqnarray*}}
\newcommand{\eean}{\end{eqnarray*}}

\newcommand{\la}{\langle}
\newcommand{\ra}{\rangle}

\def \Xh {\hat{X}}
\def \Psih {\hat{\Psi}}
\def \Ah {\hat{A}}

\def \Dh {\hat{D}}
\def \Fh {\hat{F}}

\def\p{{\partial}}
\def\ov{\over}

\title{A note on interpreting N M2 branes}

\author{Yang Zhou\footnote{yzhou@itp.ac.cn}\\
Institute of Theoretical Physics\\
Chinese Academic of Science\\
Beijing 100190, PRC\\

Interdisciplinary Center of Theoretical Studies\\
USTC, Hefei, \\
Anhui 230026, PRC \\}

\abstract{A new world volume theory with a free scalar field was proposed for Multiple M2 branes recently. By giving
the free scalar a large VEV, we can obtain a gauge theory of multiple D2 branes with a coupling constant proportional
to the VEV. It is pointed out by Banerjee and Sen that we can get D2 branes by choosing a compactification in the
$\theta$ direction, with radial position having the same value of the VEV. Following this work, we try to describe the
multiple M2 branes in $\mathbb{R}^8/\mathbb{Z}_K$ and give a simple proving of $ g_{\textit{YM}}\propto {1\ov
\sqrt{K}}$ in the compactification. We try to interpret $T^0$ and $T^{-1}$ as the radial coordinate operator and
momentum operator and give some evidences in geometry and quantum mechanics.}


\begin{document}





\date{}





\section{Introduction}

Many investigations of multiple M2 branes have been done recently.
Since the original 2+1 dimensional supersymmetic field theory with 3
bracket relation~\cite{Awata:1999dz} was proposed by Bagger, Lambert
and Gustavssonin in the
works~\cite{0611108,0709.1260,0711.0955,0712.3738,0802.3456}, many
efforts have been made to try to explain it as the theory for
multiple M2 branes. The recent progress is shown in many works,
~\cite{0803.3218}{ }{\textrm{to}}~\cite{PPCS}.

In the very recent works~\cite{0805.1012,0805.1087,0805.1202}, a
deformed algebra from the general Lie algebra was found to satisfy
the Lie 3 algebra. There are two additional generators $T^-$ and
$T^0$ on the general Lie algebra. The three product is defined as
\be\begin{split}\label{LA}
 [T^{-1},T^a,T^b]&=0\;;\\
 [T^0,T^i,T^j]&=f^{ij}{}_kT^k\;;\\
 [T^i,T^j,T^k]&=f^{ijk}T^{-1}\;.
 \end{split}
\ee where, $f^{ijk}=f^{ij}{}_lh^{lk}$ and the metric is defined as
\be
\begin{split}\label{LM}
&\la T^{-1}, T^{-1} \ra = 0, \qquad \la T^{-1}, T^0 \ra = -1, \qquad
\la T^0, T^0 \ra = 0,\\
&\la T^{-1}, T^i \ra = 0, \qquad\la T^0, T^i \ra = 0, \qquad\la T^i,
T^j \ra = h^{ij}.
\end{split} \ee
The two additional generators are very special. Take $T^{-1}$ for
example, it has a vanishing triple product with everything, which
implies $f^{(-1)ab}{}_c=0$. While $T^{0}$ can never be generated
from the triple product. It means $f^{abc}{}_0=0$. Here, we call
this Lie-3 algebra with two special generators Lorentz 3 algebra.

In the work~\cite{Banerjee:2008pd}, it is pointed out that, the
process of M2 to D2 can be clear if we choose to compactify the
$\theta$ direction. But the compact radius varies as a function of
the initial position of the M2 branes. By the duality between
M-theory and type IIA string theory, we can locally regard this
theory of M2 branes as a type IIA string theory with slowly varying
background. Along this way, we want to uncover more information
about  multiple M2 branes.

In this paper, we try to get close to the above new model for the
multiple M2 branes and give some explanations in 11D geometry. We
consider that the $r-\theta$ plane which was considered in Banerjee
and Sen's work corresponds to the level $K=1$, at which the multiple
M2 branes stay in flat 11D spacetime. For $K>1$, we see that the
flat plane becomes a cone in $r$ and $\theta$ coordinates. To do the
compactification in $\theta$ direction, we need new coordinates $r'$
and $\theta'$. And we find that the result $ g_{\textit{YM}}\propto
{1\ov \sqrt{K}}$ represents the $K$ level dependence of Yang Mill
coupling constant in N M2 branes to N D2 branes, satisfying the
result which has been obtained in $\mathcal {A}_4$.

In the picture of compactification in $\theta$ direction, we easily find that the radial coordinate which controls the
Yang Mills coupling constant corresponds to the generator $T^0$. Because in the reduction of the world volume field
theory, we can also find that $X_0$ controls the Yang Mills coupling constant by a novel Higgs mechanism. We argue that
$T^{-1}$ must correspond to the radial momentum. The N M2 branes always have constant momentum in the radial direction,
which shows that they always have a free centre of mass in $r$ direction. This picture satisfies the results well in
the world volume field theory, in which $X_0$ always have a free equation of motion. We give the picture of transition
N M2 branes in $\mathbb{R}^8/\mathbb{Z}_K$ to N D2 in varying background in Section 2 and give the explanations in
detail in the following section.

\section{Geometric description of multiple M2 branes}

So far we know little information about the N M2 branes, including
the interaction between them, which is believed to be very different
from the N D2 branes. But the recent BLG model in world volume field
theory gives us many implications. A strong method is to find the
relation between N M2 brane and N D2 branes in the field theory. The
work~\cite{0803.3218} gives a novel Higgs mechanism and the
following work~\cite{0804.1256} gives a good geometric explanation,
which is more or less complex, but there remains many puzzles. A
important one is what the physical details are for the novel
transition, from the N M2 branes to N D2 branes . The
work~\cite{Banerjee:2008pd} shows that there may be a compactifation
from M theory to IIA string theory, which seems very different from
the compactifation on a circle as before.
\subsection*{Close to Lorentz 3 algebra from Lie algebra}
First, we try to get close to the multiple M2 branes in the framwork of Lorentz 3 algebra. The original Bagger-Lambert
action of 2+1 supersymmetric gauge field can be expanded as follow ~\cite{0805.1202} \bea\begin{split}
X^I &\equiv X^I_a T^a = X^I_0 T^0 + X^I_{-1} T^{-1} + \hat{X}^I,\\
\Psi
&\equiv \Psi_a T^a = \Psi_0 T^0 + \Psi_{-1} T^{-1} + \hat{\Psi}, \\
A_{\mu} &\equiv A_{\mu a b} T^a
\otimes T^b \\
&= T^{-1} \otimes A_{\mu(-1)} - A_{\mu(-1)} \otimes T^{-1} +
T^0\otimes\hat{A}_{\mu} - \hat{A}_{\mu}\otimes T^0 + A_{\mu ij} T^i
\otimes T^j\;, \end{split}\eea where \bea \hat{X} \equiv X_i T^i,
\qquad \hat{\Psi} \equiv \Psi_i T^i,  \qquad A_{\mu(-1)} \equiv
A_{\mu (-1) a} T^a, \qquad \hat{A}_{\mu} \equiv 2 A_{\mu 0i} T^i.
\eea Thus the Bagger-Lambert Lagrangian can be written as \bea
\begin{split}
\mathcal{L} &= \langle - \frac{1}{2} (\hat{D}_{\mu}\hat{X}^I -
A'_{\mu} X^I_0)^2 + \frac{i}{4} \bar{\hat{\Psi}} \Gamma^{\mu}
\hat{D}_{\mu}\hat{\Psi} + \frac{i}{4} \bar{\Psi}_0 \Gamma^{\mu}
A'_{\mu}\hat{\Psi}
+ \frac{1}{4} (X_0^K)^2 [\hat{X}^I, \hat{X}^J]^2 \; \\
&- \frac{1}{2} (X_0^I[\hat{X}^I, \hat{X}^J])^2 + \frac{1}{2}
\epsilon^{\mu\nu\lambda} \hat{F}_{\mu\nu} A'_{\lambda}\rangle
 + \mathcal{L}_{\rm gh}\;,
\end{split}
\eea where \bea \mathcal{L}_{\rm gh} &\equiv& - \left\la \p_{\mu}
X^I_0 A'_{\mu} \hat{X}^I + (\p_{\mu} X^I_0)(\p_{\mu} X^I_{-1}) -
\frac{i}{2} \bar{\Psi}_{-1} \Gamma^{\mu}\p_{\mu} \Psi_0 \right\ra,
\eea is a ghost term for the non-positive metric, and
\bea\begin{split} \Dh_{\mu} \Xh^I \equiv \p_{\mu} \Xh^I -
[\Ah_{\mu}, \Xh^I]\;, \qquad \Dh_{\mu} \Psih \equiv \p_{\mu} \Psih -
[\Ah_{\mu}, \Psih]\;, \\ \Fh_{\mu\nu} \equiv \p_{\mu}\Ah_{\nu} -
\p_{\nu}\Ah_{\mu} - [\Ah_{\mu}, \Ah_{\nu}]\;, \qquad A'_\mu=A_{\mu
ij}f^{ij}{}_kT^k\;.
\end{split}\eea

Note that $X_{-1}$ and $\Psi_{-1}$ are Lagrange multipliers in $\mathcal{L}_{\rm gh}$ and the corresponding equation of
motions are \be \p_\mu\p^\mu X_0=0, \qquad \Gamma^\mu\p_\mu \Psi_0=0\;, \ee which means $X_{0}$ and $\Psi_{0}$ are free
fields. If we fix \be
 X^{10}_0=\nu\;, \quad \Psi_0=0\;
\ee without breaking SUSY and gauge symmetry and integrate out the
field $A'_\mu$, we can remove the ghost term $\mathcal{L}_{\rm gh}$
and obtain the effective Lagrangian \be
 \mathcal{L}_{\mbox{\small eff}} =
-\frac{1}{2}(\Dh_{\mu} \Xh^A)^2 + \frac{1}{4} v^2 [\Xh^A, \Xh^B]^2 +
\frac{i}{4} \bar{\Psih} \Gamma^{\mu}\Dh_{\mu} \Psih - \frac{1}{4
v^2} \Fh^2_{\mu\nu}, \ee where $A, B = 3, \cdots, 9$. This is the
low energy effective Lagrangian for N D2 branes.

In the work~\cite{Banerjee:2008pd}, it is pointed out that, the
process of M2 to D2 can be clear if we choose to compactify the
$\theta$ direction. But the compact radius varies as a function of
the initial position of the M2 branes. By the duality between
M-theory and type IIA string theory, we can locally regard this
theory of M2 branes as a type IIA string theory with slowly varying
background. Along this way, we want to uncover more information
about M2 to D2.

\subsection*{N M2 branes in $\mathbb{R}^8/\mathbb{Z}_K$ to N D2 in varying background}

The BLG theory based on $\mathcal {A}_4$ algebra can be written as
an ${\cal N}=8$ superconformal $SU(2) \times SU(2)$ Chern-Simons
gauge theory \cite{0803.3803}. Recently, one interesting theory was
proposed by Aharony, Bergman, Jafferis and Maldacena to find a
similar gauge theory to describe N M2 branes \cite{Aharony:2008ug}.
The former Chern-Simons matter theory
\cite{Schwarz:2004yj,Gaiotto:2007xh} was generalized to ${\cal N}=6$
superconformal form. The level of the Chern-Simons gauge theories is
$(k,-k)$ and the theory is conjectured to
 describe the low energy limit of $N$ M2-branes
 probing a $\mathbb{C}^4/\mathbb{Z}_K$. Hence at large $N$, it is
dual to the M-theory on $AdS_4 \times S^7/{\mathbb{Z}_K}$. In the
case of $SU(2) \times SU(2)$ gauge group, the theory is the same as
the Bagger-Lambert theory based on the ${\cal A}_4$ 3
algebra\cite{0803.3803}. In fact, an explanation was given recently
from ABJM to Lorentz 3 algebra in field theory~\cite{Honma:2008jd}
by scaling method.

Now we give an interpretation for N M2 branes in $\mathbb{R}^8/\mathbb{Z}_K$ to N D2 in varying background. Note that
we do not emphasize the transition in field theory but emphasize the transition in geometric picture. As proposed
in~\cite{Banerjee:2008pd}, we split the eight transverse coordinates of M2 branes into two sets, two described using
polar coordinate system and the rest of the six described using the Cartesian coordinates. We regard the azimuthal
angle as a compact coordinate and M2 branes can be regarded as D2 branes moving in a type IIA background whose dilation
and the metric vary as a function of the radial coordinate. A simple analysis shows that the effective Yang-Mills
coupling constant on N D2 branes is just the radial position. A transverse coordinate of branes is always considered as
a scalar field in the world-volume theory. From the experience of the transition from M2 to D2 in Lorentz 3 algebra, we
know that the VEV of a scalar controls the D2-brane coupling constant. We can get free equation of motion for the
radial coordinate, which satisfies the analysis in Lorentz 3 algebra.

But there remain a lot of questions. An important one is how to
describe M2 branes in $\mathbb{R}^8/\mathbb{Z}_K$ and how they
become D2 branes. We shall now proceed to describe our analysis. The
eleven dimensional flat metric is given by

\be ds_{11}^2 = \eta_{\mu\nu} dx^\mu dx^\nu + (dr^2 + r^2 d\theta^2)
+ dy^m dy^m\;,\ee where $x^\mu(\mu=0,1,2)$ are the world-volume
coordinates of the M2 branes, and $(r,\theta)$ denote the two
transverse directions of M2 brane and $y^m(m=3,4,5,6,7,8)$ are the
rest of the transverse directions. We regard the coordinate $\theta$
as the compact direction, thus the type IIA string theory will be
the weak coupling limit localized in $(\vec x, r, \vec y)$. By the
standard rule from M theory to type IIA theory, we can write the
background in 10D type IIA string theory:

\be e^{2\phi} = r^3, \qquad ds_{IIA}^2= r ( \eta_{\mu\nu} dx^\mu
dx^\nu + dr^2 + dy^m dy^m)\equiv (g_{\mu\nu}dx^\mu dx^\nu + G_{rr}
dr^2 + G_{mn} dy^m dy^n) \, . \ee In this framework, N D2 branes
localized at $(r,\vec y)$ come from N M2 branes at $(r,\theta, \vec
y)$. One interesting thing is to analyze the $SU(N)$ gauge field on
the N D2 branes after the compactification. The gauge field part in
the D2-brane world volume action is \be \label{e3} -{1\over 4} \,
\int d^3 x \, e^{-\phi} \sqrt{-\det g} \, g^{\mu\rho} \,
g^{\nu\sigma} \, Tr\left[F_{\mu\nu} F_{\rho\sigma}\right] = -{1\over
4r^{2}} \, \int d^3 x \,  \, \eta^{\mu\rho} \, \eta^{\nu\sigma} \,
Tr\left[F_{\mu\nu} F_{\rho\sigma}\right]\, . \ee Thus the effective
Yang Mills coupling constant is

\be g_{YM} = r\, . \ee

However, it was pointed out that N M2 branes in flat 11D spacetime
just correspond to gauge theory with $K=1$, where $K$ is the
Chern-Simons Level. The above picture gives a good description for
M2 branes in flat 11D spacetime, and the moduli space is
$\mathbb{R}^8$ in the world volume point of view. Now we consider
the N M2 branes in $\mathbb{R}^8/\mathbb{Z}_K$, and we must make the
moduli space smaller. While in the 11 D spacetime, we realize this
by making the regime of the $\theta$ coordinate becomes
$0\leq\theta\leq2\pi/K$, thus the original $r-\theta$ plane becomes
a cone. Or in another way, to do the compactification by the
standard rule like in Banerjee and Sen's work, keeping the volume of
moduli space (the area of $r-\theta$ plane) invariant, we can still
write the 11 D metric in another set of
coordinates\footnote{Actually, by the definition of
$\mathbb{R}^8/\mathbb{Z}_K$, we should also consider the other 6
transverse coordinates all together. We do not change them here and
they do not affect the discussion here.}

\be ds_{11}^2 = \eta_{\mu\nu} dx^\mu dx^\nu + (dr'^2 + r'^2
d\theta'^2) + dy^m dy^m\;,\ee where \be
 \theta'=K\theta\quad, r'=r/\sqrt{K}\;.
\ee By the standard rule, we can write down the type IIA dilaton
$\phi$ and the metric which comes from the 11D cone spacetime

 \be e^{2\phi} =
r'^3, \qquad ds_{IIA}^2= r' ( \eta_{\mu\nu} dx^\mu dx^\nu + dr'^2 +
dy^m dy^m)\, . \ee Then the part of D2 branes world volume action
containing the SU(N) gauge field is

\be -{1\over 4r'^{2}} \, \int d^3 x \,  \, \eta^{\mu\rho} \,
\eta^{\nu\sigma} \, Tr\left[F_{\mu\nu} F_{\rho\sigma}\right]\, . \ee
The effective Yang Mills coupling constant from the 11D cone
spacetime takes the form \be g'_{YM}=r'={g_{YM}\ov \sqrt{K}}\;,\ee
where $g_{YM}$ is the coupling constant with $K=1$.

By setting $K\rightarrow +\infty$, the effective Yang Mills coupling
$g_{YM}\rightarrow 0$ with finite $r$. Since the regime
$0\leq\theta\leq2\pi/K$ is very small, it means that the cone which
contains N M2 branes is very sharp and the radium of local circle is
very small at finite $r$. In this case, we can see the 11D spacetime
is almost compactified. And we can effectively treat the M theory as
type IIA string theory. M2 branes become D2 branes locally, and this
compactification is similar to the $S^1$ compactification locally.

This geometric interpretation of the $K$ dependence of $g_{YM}$ which can be obtained in field theory in $\mathcal
{A}_4$ gives the hints that this is a good picture of M2 branes for Lorentz 3 Algebra. Another evidence is analyzing
the scalar field $R$ of the $r$ coordinate~\cite{Banerjee:2008pd}. The equation of motion of $R$ in the type IIA theory
background is \be
 \eta^{\mu\nu} \p_\mu \p_\nu R = 0\;,
\ee which is a free equation of motion. Comparing with the $X_0^I$
which also has a free equation of motion in Lorentz 3 Algebra
theory, we can identify one of them with the other. Actually, by
giving the $X_0^{10}$ a VEV $\nu$, the effective Yang Mills coupling
is also controlled by $\nu$.

\section{M2 from D2}
We know $U(N)$ gauge theory with SUSY is the low-energy effective theory of open strings on the $D_p$ branes. In 10D
string theory, there are 9-p scalar fields $X^i$ in the adjoint representation of $U(N)$. This theory is a Lagrangian
description of the field associated with coincident $D_p$ branes. Now the question is whether it is possible to give a
similar description for M2 branes, which are the brane objects in 11D supergravity.
Lorentzian 3-algebra is a good candidate.

We can look at the Lorentz 3-Algebra in another
way~\cite{Honma:2008jd}. We rewrite (\ref{LA}) 
by defining \be \tilde{T}^{bc}X= [T^b,T^c,X]\;,\ee an using the
fundamental identity and (\ref{LA}), we get
\begin{eqnarray}
\tilde{T}^{-1a}X=0,\ \
 [\tilde{T}^{0i}, \tilde{T}^{0j}]X = i f^{ij}_k \tilde{T}^{0k}X,
\ \
 [\tilde{T}^{0i}, S^j]X = i f^{ij}_k S^kX, \ \
 [S^i, S^j]X=0.
 \label{BLalgebra}
\end{eqnarray}
where $S^i=f^i_{jk} \tilde{T}^{jk}$. This algebra shows that the
Lorentz 3 Algebra just contains some rotations and translations. It
implies that we can find some physical explanation for this algebra.
Before discussing this problem, we need a framework.

 As pointed out
in the work~\cite{Banerjee:2008pd}, if we compactify M theory on a
cone, then the the compactification radium depending on the radial
position of N M2 branes will give the coupling constant in SYM of N
D2 branes and satisfy the equation of motion of free scalar. It
implies that the Lorentz 3 algebra will give more information from N
M2 branes to D2 branes in Banerjee and Sen's geometry framework.

This framework is also good for us to investigate M2 branes
themselves. Because there is a special coordinate $r$ even before
compactification. This transverse coordinate is believed to be a
free scalar in Lorentz 3 algebra. The free equation of motion comes
from the Lagrange multiplier. The coupling constant of D2 branes
becomes smaller when the radial position of M2 branes decreases. It
means that small radial position makes M2 branes become similar to
D2 branes. For $r\rightarrow0$, N M2 branes look like the N D2
branes and the radius of the circle is very small and not apparent
to be found. If we want to find the original configuration in M2
branes for the IIA fundamental string, we know it is a M2 brane
wrapped on a circle. We believe that when we compactify the N M2
branes, the picture of interaction of branes is similar to the D2
branes.


We can show the above picture of M2 branes as below. We choose a
transverse plane $(r,\theta)$ transverse to N M2 branes. The flat
metric in 11D is

\be ds_{11}^2 = \eta_{\mu\nu} dx^\mu dx^\nu + (dr^2 + r^2 d\theta^2) + dy^m dy^m\, . \ee We choose static gauge and a
certain embedding method to let $x^\mu(0\leq\mu\leq2)$ be world-volume coordinates of M2-brane and $y^m(3\leq m\leq7)$
are the rest transverse coordinates. Following the proposal in work~\cite{Banerjee:2008pd}, we can interpret the
coordinate $\theta$ as the compact direction. Then N M2 branes placed at $(r,\theta,\vec y)$ can be regarded as N
coincident D2-branes placed at $(r,\vec y)$. $\theta$ denotes the position of N M2 brane before the compactification.
We now want to find some interpretations for the Lorentz 3 algebra
which gives a good framework for the transformation from N M2 branes
to D2 branes. We give our interpretation of the algebra relation in
(\ref{BLalgebra}) using the method in quantum mechanics as follow:

1. We treat all the fields as operators in quantum mechanics. All
the transverse coordinates $X^I(I=3,4,5,6,7,8,9,10)$ are the
position operators of the quantum state of whole N M2 branes system.

2. $T^a(a=-1,0,..\dim(\mathcal {G}_{SU(N)}))$ are the basic
operators. $T^{0}$ is of course the radial coordinate operator, and
$T^{-1}$ is the corresponding conjugate operator. $T^{-1}$ also can
be interpreted as momentum operator in the radial coordinate.

3. $T^i(i=1....\dim(\mathcal {G}_{SU(N)}))$ are inner generators
corresponding the inner $SU(N)$ symmetry.

4. The Lorentz 3 algebra describe N M2 branes system which always
stays in the eigenstate of $T^{-1}$. It also means that the Lorentz
3 algebra describes N M2 system with free centre of mass $r$.


With the above statements, we can understand N M2 branes in the Lorentz 3 algebra. For the fixed moment which is
decoupled with other operators and free $r$ of the system, we have

\be \tilde{T}^{-1a}=0,\quad
 [\tilde{T}^{0i}, \tilde{T}^{0j}] = i f^{ij}_k \tilde{T}^{0k}.
 \ee
The relation
\be
 [\tilde{T}^{0i}, S^j] = i f^{ij}_k S^k,\quad
 [S^i, S^j]=0
\ee can be related to \be [T^i,T^j,T^k]=f^{ijk}T^{-1}\;.\ee We know
\be
 \tilde{T}^{ij}X= [T^i,T^j,X]=f^{ijk}X_kT^{-1}\;,
\ee which shows that the 3 bracket transformation gives a
translation in $r$ direction. In fact, the fields related to
$T^{-1}$ are all Lagrange multipliers and have no dynamics. Because
N branes system has no force in the $r$ direction, the momentum is
always a constant. The gauge fields still need to be discussed.

\section{Conclusion}

It is pointed out by Banerjee and Sen that we can get D2 branes by choosing a compactification in the $\theta$
direction, with radial position having the same value of the VEV. In this letter, we try to get close to this new model
for the multiple M2 branes and give some explanations in 11D geometry. We consider that the $r-\theta$ plane which was
considered in Banerjee and Sen's work corresponds to the level $K=1$, at which the multiple M2 branes stay in flat 11D
spacetime. For $K>1$, we see that the flat plane becomes a cone in $r$ and $\theta$ coordinates. And we find that the
result $ g_{\textit{YM}}\propto {1\ov \sqrt{K}}$ represents the $K$ level dependence of Yang Mill coupling constant in
N M2 branes to N D2 branes, satisfying the result which has been obtained in $\mathcal {A}_4$.

In the picture of compactification in $\theta$ direction, we easily find that the radial coordinate which controls the
Yang Mills coupling constant corresponds to the generator $T^0$. Because in the reduction of the world volume field
theory, we can also find that $X_0$ controls the Yang Mills coupling constant by a novel Higgs mechanism. We argue that
$T^{-1}$ must correspond to the radial momentum. The N M2 branes always have constant momentum in the radial direction,
which shows that they always have a free centre of mass in $r$ direction. This picture satisfies the results well in
the world volume field theory, in which $X_0$ always have a free equation of motion.
\section*{Related discussion}
In the general $S^1$ compactification framework, fundamental strings in 10D IIA theory are the wrapped M2 branes on the
circle in 11D M theory. And the D2 branes are the M2 branes extended in the transverse directions to the circle.
To describe the low energy effective theory of multiple coincident D branes theory, we can use open string picture. In
multiple M2 branes, things are different. There are no objects like open strings which can transport the interaction
between the branes. In fact, D2 branes are not so heavy as in perturbative IIA string theory when the string coupling
constant becomes very large. Thus they can not keep static easily when we consider their interaction and fluctuation.
It is difficult to describe the world volume field theory. From the recent BL action we know that the transverse scalar
excitations have kinetic terms and an interaction potential, while the gauge field has no kinetic term but only
topological term. We can describe the scalars on M2 branes like ones on D2 branes more or less, only with different
gauge transformation.
Our proposal relates the inner gauge symmetry and spacetime symmetry
and try to give some interpretations for the M2 branes. Before
compactification, N M2 branes are free in the $r$ direction and have
a constant momentum in that direction. Through compactifying the
$\theta$ direction, we have a type IIA string theory locally. When
$K=1$, the flat 11D background is a plane, while it will be a cone
of ($r,\theta$) coordinates at a large $K$.

After the above discussion, we find that multiple M2 branes are very different from the usual D2 branes. The special
free coordinate $r$ makes M2 branes almost cover D2 branes with different coupling constants in 11D M theory. Due to
the big coupling constant, we guess that the additional symmetries originate from the quantum properties of the whole
M2 branes system. A further clear explanation and a more apparent expression for our picture are under consideration.

After this work has been almost finished, Bagger and Lambert give a
work to try to construct a new algebra in which the triple product
is only antisymmetric in two indices. They claimed that they can obtain
the ${\cal N}=6$ theories that have been recently proposed as models for M2 branes
 in an $\mathbb{R}^8/\mathbb{Z}_K$ orbifold background~\cite{Aharony:2008ug}. This
 modified algebra can give some evidences to support our conjecture.
 In our statements, some special operators like $T^{-1}$ and $T^0$ do not play the
 same roles in the triple product. We can also write the Lorentz 3
 algebra using the modified triple product and find some interesting
 properties. The modified triple product is natural in our
 conjecture, and the most important reason is the space time symmetries of the
 whole N M2 branes system are different from the inner gauge
 symmetries.




\section*{Acknowledgements} The
author acknowledges helpful discussions with Chaojun Feng and
Zhaolong Wang and thanks Miao Li and Yi Wang for reading this paper.

\end{document}